\documentclass[sigplan, screen]{acmart} 
\usepackage[utf8]{inputenc}
\usepackage{booktabs}
\usepackage{amsmath}
\usepackage{fancyvrb}
\usepackage{subcaption}
\usepackage{algorithm, algorithmicx}
\usepackage[noend]{algpseudocode}
\usepackage{todonotes}

\newcommand{\expD}{\mathcal{X}^{N}_d}
\newcommand{\uniD}{\mathcal{U}^{N}_d}

\title{Automatic Synthesis of Experiment Designs from Probabilistic Environment Specifications}
\author{Craig Innes}
\affiliation{%
\institution{University of Edinburgh}
}
\email{craig.innes@ed.ac.uk}

\author{Yordan Hristov}
\email{yordan.hristov@ed.ac.uk}
\affiliation{%
\institution{University of Edinburgh}
}

\author{Georgios Kamaras}
\email{g.kamaras-1@sms.ed.ac.uk}
\affiliation{%
\institution{University of Edinburgh}
}

\author{Subramanian Ramamoorthy}
\email{s.ramamoorthy@ed.ac.uk}
\affiliation{%
\institution{University of Edinburgh}
}

\keywords{computational geometry, design of experiments, probabilistic specification, verification, robotics}

\setcopyright{none}
\copyrightyear{2021}
\acmYear{2021}
\acmDOI{}

\acmConference[SYNT '21]{}{July 19th 2021}{Los Angeles, CA}
\acmISBN{}

\begin{document}


\maketitle

\section{Introduction}



Imagine you have designed an algorithm for a tabletop robotic arm. You had a \emph{task specification} (e.g., move an object from the in-tray to an out-tray). You likely also had an \emph{environment specification}---a description of the range of environments the robot should perform successfully in (e.g., ``robot at table's centre, two trays at any position on the table, object weighs 500-1000g and always starts in the in-tray''). Across all potential variations in position, size, and weight, you now wish to assess your algorithm's performance (e.g., with an \textsc{stl} robustness metric based on your task specification \cite{mehdipour2019arithmetic,varnai2020robustness,haghighi2019control}).

\begin{figure}
\centering
\includegraphics[width=\linewidth]{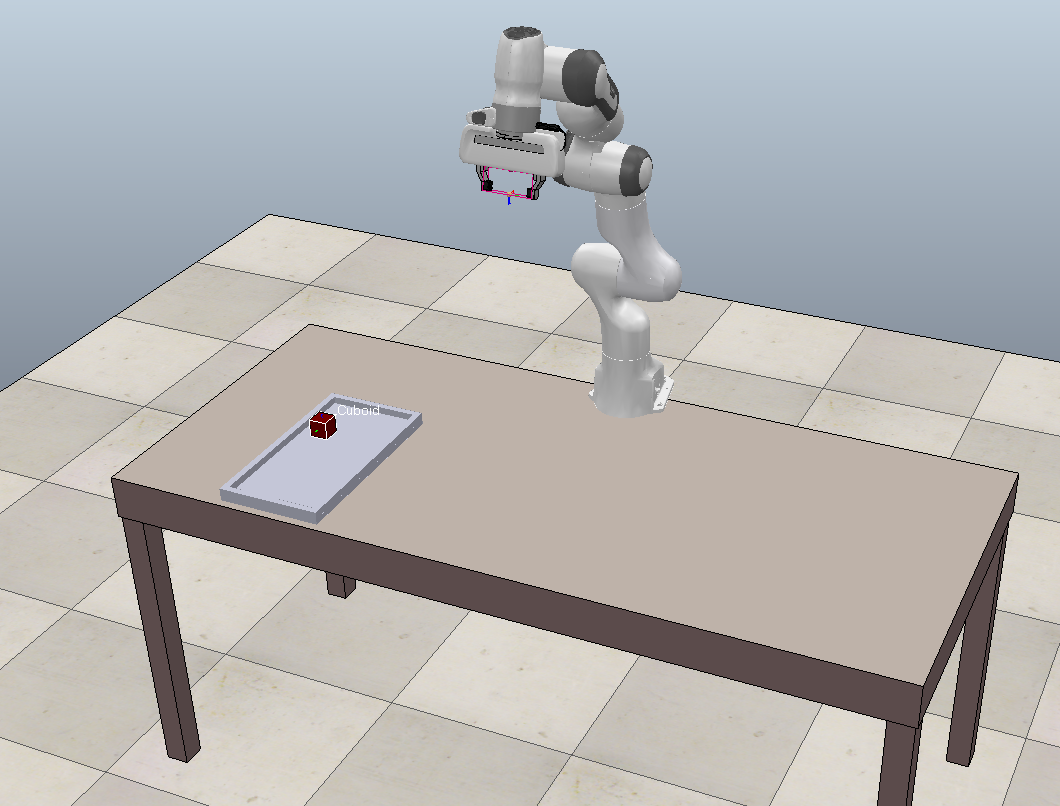}
\caption{Tabletop scene generated from \textsc{prs} spec (\ref{fig:experiment:tray-cube-table}).}
\end{figure}

Checking all possible environments is usually impossible: one robotic experiment, even simulated, takes non-trivial time; We cannot check all values for continuous variables like object position without first discretizing them; assessing all interaction effects between parameters requires an exponential combination of values. Realistically, we have time to try only a finite subset of the possible environments. 

Given a budget of $N$ experiments, how can we pick the best subset of possible values which will give us best coverage of the possible range of environments? We could draw $N$ environment configurations at random; Frameworks like Scenic \cite{fremont2018scenic} or ProbRobScene \cite{innes2020probrobscene}  provide a way to automatically generate samplers for declarative descriptions of an environment specification. However, there is no guarantee a random sample of configurations will best represent the range of parameter values, nor interactions between parameters---random samples tend to have \emph{high discrepancy} \cite{garud2017design}.

To generate \emph{low-discrepancy} (\emph{uniform}) sample sets, there exist established methods in \emph{experiment design}. Such methods generate sample sets (\emph{designs}) on the $d$-dimensional unit-hypercube. However, in robot experiments like the one above, the parameter space is not a hypercube, but a constrained geometric region. Generating low-discrepancy samplings over such regions often requires knowledge of special properties of the space \cite{tian2009uniform}. Further, large environment specifications may require a design over not just one, but multiple (possibly dependent) regions. Typical users lack the expertise to construct such complex designs, despite being able to express their environment specification declaratively.

This paper augments the probabilistic programming language ProbRobScene (\textsc{prs}) \cite{innes2020probrobscene} with a module for automatically synthesizing uniform designs. It builds on methods for constructing uniform designs over irregular domains \cite{zhang2020construction}, dependency resolution \cite{fremont2018scenic}, and algorithms for convex geometry \cite{innes2020probrobscene}. We also include a case-study which uses snippets of an environment specification from a tabletop scenario. Our experiments show that our designs achieve lower discrepancy than random-sampling for a range of $N$. We also show how the computation time varies with $N$ and $d$.

\section{Uniform Designs and Environment Specs}

We first review uniform designs on a unit hypercube, and the \textsc{prs} language. We explain our method for synthesizing designs from arbitrary \textsc{prs} specifications in the next section.

\subsection{Uniform Designs on the Unit HyperCube}

Say we have a budget for $N$ experiments, $\expD = \{ x^{(i)} \in D | i = 0 \dots N \}$ where our domain $D$ is defined by $d$ parameters in the unit range (i.e., $D = [0,1]^d$). We wish to pick an $\expD$ with \emph{low discrepancy} over $D$. \emph{Discrepancy} measures how much $\expD$ differs from the ideal uniform design---one where, in every subspace $\Delta D \subset [0, 1]^d$, the proportion of points in $\Delta D$ is proportional to the size of $\Delta D$ \cite{garud2017design}.

\begin{figure}
    \centering
    \includegraphics[width=\linewidth]{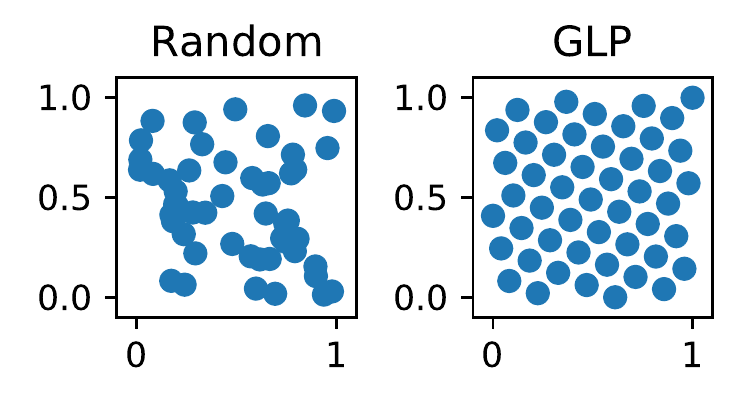}
    \caption{Visualization of $N=50$ samples in  $[0,1]^2$ for Random vs \textsc{GLP}}
    \label{fig:hypercube-comparison}
\end{figure}

There exist standard methods to generate low discrepancy unit hyper-cube designs. For example the \emph{Good Lattice Point Method} (\textsc{glp}) \cite{zaremba1966good} leverages the relative primes of $N$ to produce reliably low-discrepancy designs. Figure (\ref{fig:hypercube-comparison}) shows random sampling versus \textsc{glp} on the unit-square.

\subsection{\textsc{prs} Specs as Dependent Convex Regions}

\begin{figure}
\begin{subfigure}{\linewidth}
    \centering
    \begin{LVerbatim}[fontsize=\footnotesize]
t = Table on V3D(0,0,0)
r1 = Robot on (top back t)
Cube completely on t, with mass (500, 1000)
    \end{LVerbatim}
    \caption{\emph{``Cube-Table''}}
    \label{fig:lang-examples:single}
\end{subfigure}

\begin{subfigure}{\linewidth}
    \centering
    \begin{LVerbatim}[fontsize=\footnotesize]
t = Table on V3D(0,0,0)
r1 = Robot on (top back t)
tr_1 = Tray completely on t, ahead of r1, left of t
Cube completely on tr_1
    \end{LVerbatim}
    \caption{\emph{``Cube-Tray-Table''}}
    \label{fig:lang-examples:dependent}
\end{subfigure}
\caption{Examples of \textsc{prs} environment specifications}
\label{fig:lang-examples}
\end{figure}

Probabilistic programming languages can be used declaratively specify environment variations in domains such as autonomous vehicles \cite{fremont2018scenic}, image generation \cite{kulkarni2015picture}, and procedural modelling \cite{ritchiequicksand}. Here, we focus on \textsc{prs}, which is well suited to setting up tabletop manipulation environments. Figure (\ref{fig:lang-examples}) shows two examples of \textsc{prs} environment specifications. \textsc{prs} leverages the \emph{specifier syntax} (from Scenic \cite{fremont2018scenic}) to let the user describe the value ranges for object properties in three ways. First, they can provide a scalar (e.g., set the z-coordinate of \texttt{t} to 0). Second, they can provide a uniform range (e.g., setting \texttt{mass} of \texttt{Cube} to between 500 and 1000). Third, they can define range of values for a property via a \emph{geometric relation} (e.g., \texttt{tr\_1} is \emph{ahead of} \texttt{r1}). 

\textsc{prs} generates samples from this specification in several steps. First, \textsc{prs} converts each geometric relation into a convex-set. Since some properties are defined as the combination of multiple relations, this may also involve using algorithms for intersecting convex polytopes. For example, in (\ref{fig:lang-examples:dependent}), the tray position is intersection of the table surface, and the two half-spaces in front of the robot and to the left of the table centre. Next, \textsc{prs} determines which order to evaluate the specifiers, as some depend on each other. Again in figure (\ref{fig:lang-examples:dependent}): \texttt{Cube}'s position depends on the size and position of \texttt{tr\_1}, which depends on the size and position of \texttt{t}. Finally, once the dependency order and sampling regions are defined, \textsc{prs} draws samples from the constrained space using hit-and-run---a uniform sampling method for convex polytopes \cite{Zabinsky2013}.

Using this framework, users can declaratively specify the range of environments they expect their robot to operate in, and automatically generate samples which conform to this specification. However, as discussed in the previous section, generating arbitrary samples from a space tends to result in a high-discrepancy designs, and is therefore typically a poor representation of the performance of the system. We now show how we extend this system such to automatically synthesize uniform designs.

\section{Design Synthesis from \textsc{prs} Specifications}

Recall our setup---we have $N$ experiments to generate a design $\expD$ with low discrepancy over $D$. However, now $D$ is not a unit hypercube $[0,1]^d$, but a combination of dependent convex regions. Can we generate a uniform design over this new domain by building on our ability to generate a unit-hypercube design $\uniD = \{ u^{(i)} \in [0,1]^d \mid i = 0 \dots N \}$? Recent work in transforming uniform designs gives us an insight: Let $x \in \mathcal{D} \subseteq \mathbb{R}^d$ be a random vector with density:

\begin{equation}
    f(x_1, \dots, x_d) = f_1(x_1)f_{2|1}(x_2 | x_1) \dots f_{d|1\dots}(x_d | x_1, \dots)
    \label{eqn:bayes-rule}
\end{equation}

If we can write the cumulative distribution functions (\textsc{cdf}s) for each factor of (\ref{eqn:bayes-rule}) (denoted $F_{1}, F_{2|1} \dots F_{d, 1\dots}$), then we can map $\mathcal{D}$ to $[0,1]^d$:

\begin{equation}
\begin{cases}
    \quad u_1 = F_1(x_1) \\
    \quad u_j = F_{j | 1 \dots j-1}(x_j | x_1, \dots, x_{j-1}),\ j=2,\dots, d
\end{cases}
\label{eqn:cdfs}
\end{equation}

Given (\ref{eqn:cdfs}), we can convert $\uniD$ to a uniform design $\expD$ over $D$ using the \emph{Inverse Rosenblatt Transformation} \cite{zhang2020construction}: 

\begin{equation}
\begin{cases}
    \quad x^{(i)}_1 = F_{1}^{-1}(u^{(i)}_1) \\
    \quad x^{(i)}_j = F_{j|1 \dots j-1}^{-1}(u_j^{(i)} \mid x^{(i)}_1 \dots x^{(i)}_{j-1}),\  j = 2,\dots,d \\
\end{cases}
\label{eqn:inv-cdf}
\end{equation}

To construct the \textsc{cdf}s for our specification, recall that each geometric relation in \textsc{prs} can be represented by a convex-set. Each region can therefore be written compactly as: 

\begin{equation}
    R_{x} = \{x \in \mathbb{R}^k \mid Ax + b \leq 0 \} \\
\label{eqn:regions}
\end{equation}

$A \in \mathbb{R}^{m \times k}$ and $b \in \mathcal{R}^m$ are the coefficients of $m$ convex inequalities. Using (\ref{eqn:regions}) we can construct a \textsc{cdf} for a uniform distribution over a convex region by integrating over $R_x$. For a 3-dimensional convex polyhedron, $F_1$ would be given by: 

\begin{equation}
    F_1(x^{(i)}_1) = \iiint^{x^{(i)}_1} \frac{1}{V(R_x)}\ \partial x_1 \partial x_2 \partial x_3  \\
\label{eqn:region-cdfs}
\end{equation}

Where $V(R)$ is the volume of region $R$. The conditional \textsc{cdf}s such as $F_{2|1}(x^{(i)}_2 | x^{(i)}_1)$ can be computed similarly by first projecting $R_x$ onto the relevant hyper-plane. While we lack a closed form representation for the inverse \textsc{cdf}s required to calculate (\ref{eqn:inv-cdf}), we can approximate each transformation from $u^{(i)}$ to $x^{(i)}$ numerically using Brent's method \cite{brent2013algorithms}.

We can now generate uniform designs over convex regions. To synthesize a uniform design over a full \textsc{prs} specification, we have two further difficulties. 

First, \textsc{prs} environment specifications are actually comprised of several convex regions, whose boundaries depend on one another (e.g., recall from Figure (\ref{fig:lang-examples:dependent})--- the \texttt{Cube} position depends on \texttt{tr\_1}, which depends \texttt{Table}). We must therefore construct a tree of these dependencies, similar to the process followed when sampling. This dependency tree limits on the possible factorings of (\ref{eqn:bayes-rule}) and, by extension, the orders we can condition the \textsc{cdfs} in (\ref{eqn:cdfs}, \ref{eqn:inv-cdf}).

The second difficulty is that our design is dependent on the order we condition each \textsc{cdf} dimension --- $x^{(i)}_1 = F^{-1}_1(u^{(i)}_{1})$, $x^{(i)}_2 = F_{2 | 1}^{-1}(u^{(i)}_{2} | x^{(i)}_1)$ produces a different design from $x^{(i)}_2 = F^{-1}_2(u^{(i)}_{2})$, $x^{(i)}_1 = F_{1 | 2}^{-1}(u^{(i)}_{1} | x^{(i)}_2)$. To decide the best, we pick the $\expD$ with minimal \emph{central composite discrepancy} (\textsc{ccd}) \cite{chuang2010uniform}:

\begin{equation}
    \textsc{ccd}(\expD) = \left\{ \frac{2^{-d}}{V(\mathcal{D})} \int\limits_{\mathcal{D}}\sum^{2^d}_{k=1} \left\lvert \frac{|\expD \cap \mathcal{D}_k(z)|}{N} - \frac{V(\mathcal{D}_k(z))}{V(\mathcal{D})} \right\rvert^2 \partial z \right\}^{\frac{1}{2}}
    \label{eqn:ccd}
\end{equation}

Here, $\mathcal{D}_k(z)$ is $k$-th section if we partition $\mathcal{D}$ into $2^d$ sections, splitting across every axis from centre $z$. Intuitively, (\ref{eqn:ccd}) measures the difference between the proportion of points in each sub-region, and the sub-region's relative volume.

\begin{algorithm}
        \caption{Uniform Design from \textsc{prs} Specification}
        \label{alg:full-system}
        \begin{algorithmic}[1]
        \Function{Synthesize-Design}{\textit{prs-spec}, $N$}
            \State \textit{viable-orders} $\gets$ \Call{extract-dependencies}{\textit{prs-spec}}
            \State $\mathcal{D} \gets$ \Call{convert-to-convex-regions}{\textit{prs-spec}}
            \State $\uniD \gets$ \Call{Good-Lattice-Point}{$N$, $d$}
            \For{\textit{d-order} $\in$ \textit{viable-orders}}
                \State $\textit{designs}_i \gets$ \Call{Inverse-Rosenblatt}{$\uniD$, \textit{d-order}}
                \State $CCD_i \gets$ \Call{CCD}{$\textit{cand-designs}_i$}
            \EndFor
            \State \Return $\expD$ in $\textit{designs}$ with lowest CCD
        \EndFunction
        \end{algorithmic}
\end{algorithm}

Algorithm (\ref{alg:full-system}) summarizes the full process. First, the \textsc{prs} specification is converted into individual convex regions, and an object property dependency graph is extracted. Next, we loop through viable dimension orderings (respecting restrictions imposed by the dependency graph), generate candidate design via (\ref{eqn:inv-cdf}), then choose the design with minimal \textsc{ccd}. 

\begin{figure}
\centering
\begin{subfigure}{\linewidth}
\centering
    \caption{\emph{cube-table}}
    \includegraphics[width=0.49\textwidth]{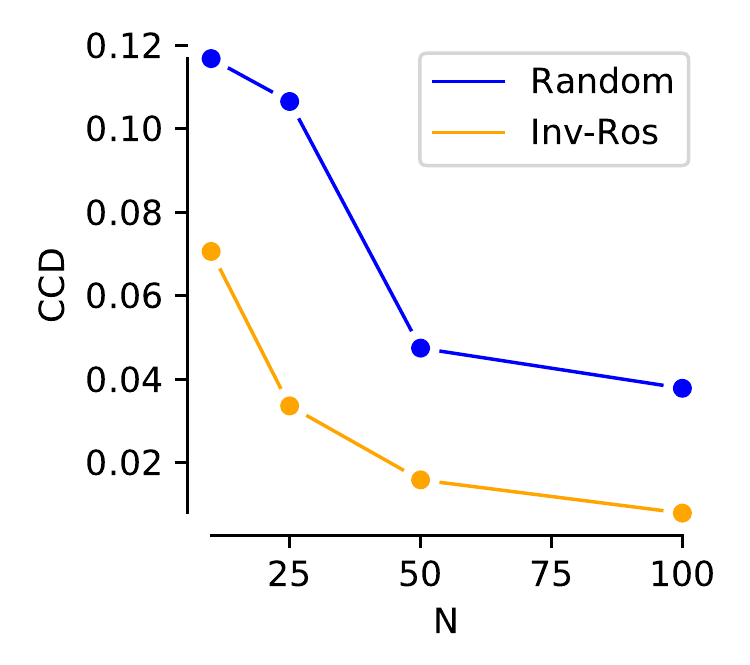}
    \includegraphics[width=0.49\textwidth]{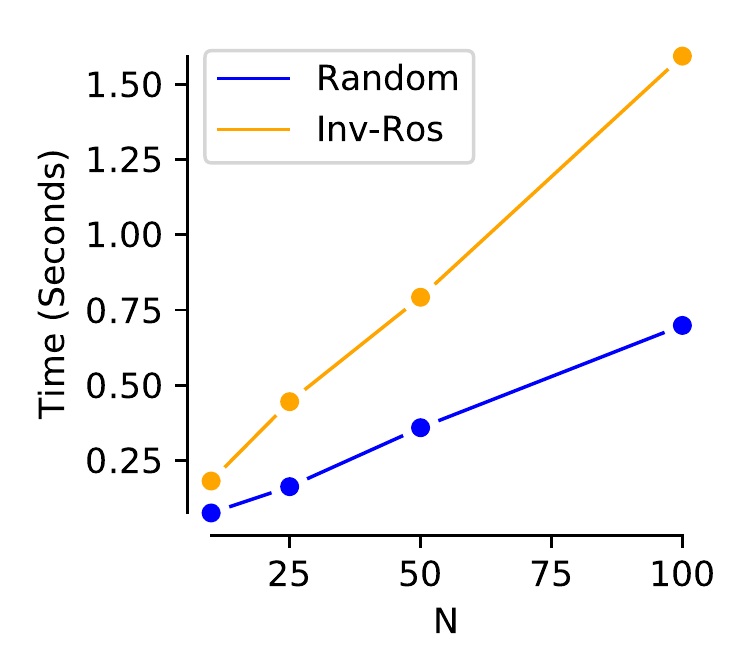}
    \label{fig:experiment:cube-table}
\end{subfigure}

\begin{subfigure}{\linewidth}
    \centering
    \caption{\emph{tray-cube-table}}
    \includegraphics[width=0.49\textwidth]{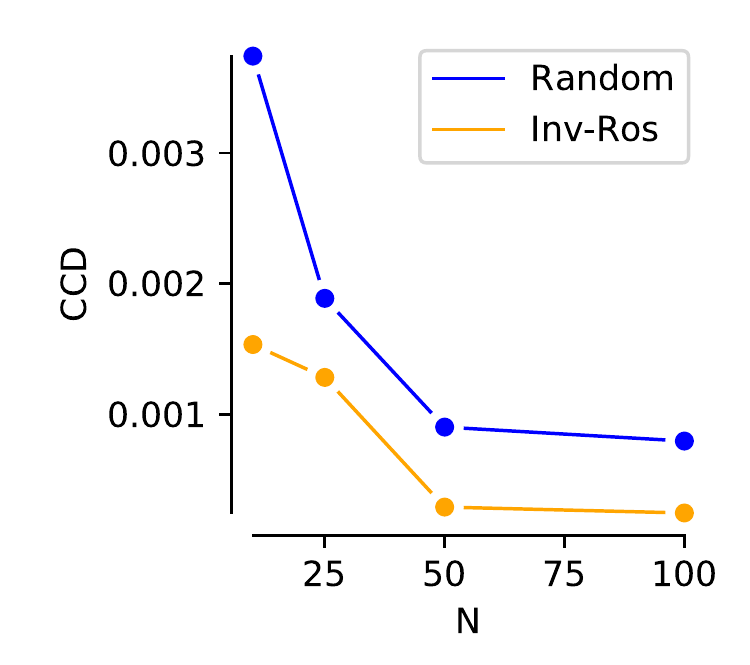}
    \includegraphics[width=0.49\textwidth]{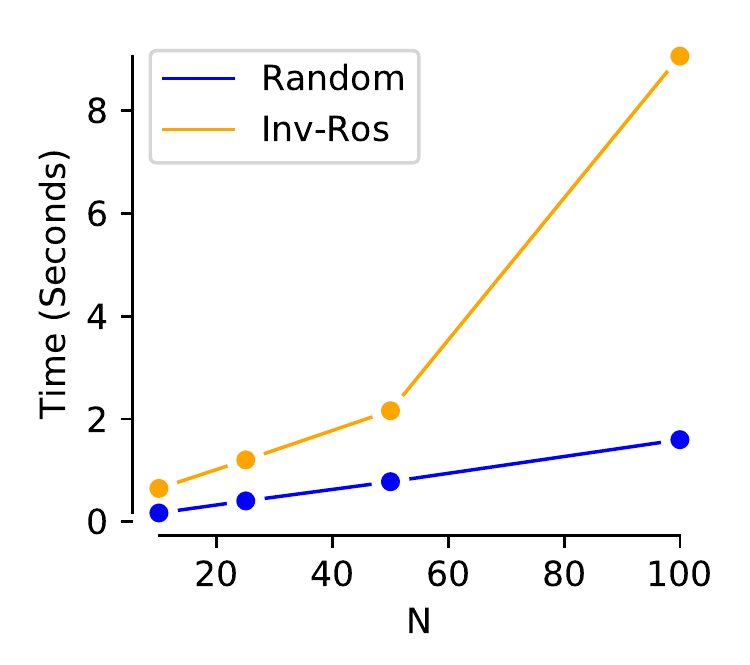}
    \label{fig:experiment:tray-cube-table}
\end{subfigure}
\caption{\textsc{ccd}s and times for $N=\{10, 25, 50, 100\}$}
\label{fig:experiment}
\end{figure}

\section{Case Study: Tabletop Setups}

This section demonstrates our augmented \textsc{prs} module on two environment specification snippets from tabletop manipulation examples (Figures (\ref{fig:lang-examples:single}) and (\ref{fig:lang-examples:dependent})). These cover the single convex region case and the dependent region case.

Figure (\ref{fig:experiment}) shows the \textsc{ccd} and computation time for our method for $N=\{10, 25, 50, 100\}$. Our system achieves reliably low-discrepancy designs at all values of $N$, with lower discrepancies as $N$ increases. Contrast this with random sampling, which has consistently higher discrepancies.

One thing to note in the change from (\ref{fig:experiment:cube-table}) to (\ref{fig:experiment:tray-cube-table}) is the change in relative computation time. While random sampling remains quick and scales linearly with $N$, our method appears to increase both in absolute time and relative growth. The reason for this is that the bulk of the computation resides in the inverse-rosenblatt transformation, and the number of transformations to consider grows factorially with $d$. In practice, the computation time taken to generate the experimental designs is usually completely dominated by the time taken to actually run the experiments. However, for complex specifications with large $d$, heuristics for searching the space of possible transformations may have to be used to preserve scalability.

\section{Related Work and Discussion}

Most existing probabilistic programming languages offer only the ability to draw Monte-Carlo samples \cite{fremont2018scenic, kulkarni2015picture, ritchiequicksand}. VerifAI \cite{dreossi2019verifai} allows users to manually configure external quasi-random samplers (e.g., Halton sampling \cite{niederreiter1992random}) which tend achieve lower discrepancies. None offer the ability to synthesize uniform designs directly from specifications.

Our module builds off a generic technique for synthesizing designs using Inverse Rosenblatt transforms \cite{zhang2020construction}, but there are also specialized techniques based on particle swarms \cite{chen2014central}, and identifying specific sub-classes of convex polyhedrons \cite{tian2009uniform}. Such specialized methods tend not to generalize to arbitrary specifications, but could be used as pruning heuristics to improve computational efficiency in larger specifications.

Our synthesis approach can be classified as \emph{system-free}, as it favours low discrepancy designs without reference to a task-specific metric. However, in the field of robot manipulation, there exist numerous metrics for evaluating task success with respect to a formal task-specification \cite{mehdipour2019arithmetic, varnai2020robustness, haghighi2019control}. One direction for future work is to use such metrics to drive \emph{adaptive experiment design}. Such methods choose samples based on uncertainty over predicting the task metric \cite{eason2014adaptive}.

\begin{acks}
This work is supported in part by funding from the Alan Turing Institute, as part of the Safe AI for Surgical Assistance project, and from EPSRC for the UKRI Research Node on Trustworthy Autonomous Systems Governance and Regulation (EP/V026607/1).
\end{acks}



\bibliographystyle{ACM-Reference-Format}
\bibliography{references}


\begin{thebibliography}{18}


\ifx \showCODEN    \undefined \def \showCODEN     #1{\unskip}     \fi
\ifx \showDOI      \undefined \def \showDOI       #1{#1}\fi
\ifx \showISBNx    \undefined \def \showISBNx     #1{\unskip}     \fi
\ifx \showISBNxiii \undefined \def \showISBNxiii  #1{\unskip}     \fi
\ifx \showISSN     \undefined \def \showISSN      #1{\unskip}     \fi
\ifx \showLCCN     \undefined \def \showLCCN      #1{\unskip}     \fi
\ifx \shownote     \undefined \def \shownote      #1{#1}          \fi
\ifx \showarticletitle \undefined \def \showarticletitle #1{#1}   \fi
\ifx \showURL      \undefined \def \showURL       {\relax}        \fi
\providecommand\bibfield[2]{#2}
\providecommand\bibinfo[2]{#2}
\providecommand\natexlab[1]{#1}
\providecommand\showeprint[2][]{arXiv:#2}

\bibitem[\protect\citeauthoryear{Brent}{Brent}{2013}]%
        {brent2013algorithms}
\bibfield{author}{\bibinfo{person}{Richard~P Brent}.}
  \bibinfo{year}{2013}\natexlab{}.
\newblock \bibinfo{booktitle}{\emph{Algorithms for minimization without
  derivatives}}.
\newblock \bibinfo{publisher}{Courier Corporation}.
\newblock


\bibitem[\protect\citeauthoryear{Chen, Shu, Hung, and Wang}{Chen
  et~al\mbox{.}}{2014}]%
        {chen2014central}
\bibfield{author}{\bibinfo{person}{Ray-Bing Chen}, \bibinfo{person}{YH Shu},
  \bibinfo{person}{Ying Hung}, {and} \bibinfo{person}{Weichung Wang}.}
  \bibinfo{year}{2014}\natexlab{}.
\newblock \showarticletitle{Central composite discrepancy-based uniform designs
  for irregular experimental regions}.
\newblock \bibinfo{journal}{\emph{Computational Statistics and Data Analysis}}
  \bibinfo{volume}{72} (\bibinfo{year}{2014}), \bibinfo{pages}{282--297}.
\newblock


\bibitem[\protect\citeauthoryear{Chuang and Hung}{Chuang and Hung}{2010}]%
        {chuang2010uniform}
\bibfield{author}{\bibinfo{person}{SC Chuang} {and} \bibinfo{person}{YC Hung}.}
  \bibinfo{year}{2010}\natexlab{}.
\newblock \showarticletitle{Uniform design over general input domains with
  applications to target region estimation in computer experiments}.
\newblock \bibinfo{journal}{\emph{Computational Statistics \& Data Analysis}}
  \bibinfo{volume}{54}, \bibinfo{number}{1} (\bibinfo{year}{2010}),
  \bibinfo{pages}{219--232}.
\newblock


\bibitem[\protect\citeauthoryear{Dreossi, Fremont, Ghosh, Kim, Ravanbakhsh,
  Vazquez-Chanlatte, and Seshia}{Dreossi et~al\mbox{.}}{2019}]%
        {dreossi2019verifai}
\bibfield{author}{\bibinfo{person}{Tommaso Dreossi}, \bibinfo{person}{Daniel~J
  Fremont}, \bibinfo{person}{Shromona Ghosh}, \bibinfo{person}{Edward Kim},
  \bibinfo{person}{Hadi Ravanbakhsh}, \bibinfo{person}{Marcell
  Vazquez-Chanlatte}, {and} \bibinfo{person}{Sanjit~A Seshia}.}
  \bibinfo{year}{2019}\natexlab{}.
\newblock \showarticletitle{Verifai: A toolkit for the formal design and
  analysis of artificial intelligence-based systems}. In
  \bibinfo{booktitle}{\emph{International Conference on Computer Aided
  Verification}}. Springer, \bibinfo{pages}{432--442}.
\newblock


\bibitem[\protect\citeauthoryear{Eason and Cremaschi}{Eason and
  Cremaschi}{2014}]%
        {eason2014adaptive}
\bibfield{author}{\bibinfo{person}{John Eason} {and} \bibinfo{person}{Selen
  Cremaschi}.} \bibinfo{year}{2014}\natexlab{}.
\newblock \showarticletitle{Adaptive sequential sampling for surrogate model
  generation with artificial neural networks}.
\newblock \bibinfo{journal}{\emph{Computers \& Chemical Engineering}}
  \bibinfo{volume}{68} (\bibinfo{year}{2014}), \bibinfo{pages}{220--232}.
\newblock


\bibitem[\protect\citeauthoryear{Fremont, Yue, Dreossi, Ghosh,
  Sangiovanni-Vincentelli, and Seshia}{Fremont et~al\mbox{.}}{2018}]%
        {fremont2018scenic}
\bibfield{author}{\bibinfo{person}{Daniel Fremont}, \bibinfo{person}{Xiangyu
  Yue}, \bibinfo{person}{Tommaso Dreossi}, \bibinfo{person}{Shromona Ghosh},
  \bibinfo{person}{Alberto~L Sangiovanni-Vincentelli}, {and}
  \bibinfo{person}{Sanjit~A Seshia}.} \bibinfo{year}{2018}\natexlab{}.
\newblock \showarticletitle{Scenic: Language-based scene generation}.
\newblock \bibinfo{journal}{\emph{arXiv preprint arXiv:1809.09310}}
  (\bibinfo{year}{2018}).
\newblock


\bibitem[\protect\citeauthoryear{Garud, Karimi, and Kraft}{Garud
  et~al\mbox{.}}{2017}]%
        {garud2017design}
\bibfield{author}{\bibinfo{person}{Sushant~S Garud},
  \bibinfo{person}{Iftekhar~A Karimi}, {and} \bibinfo{person}{Markus Kraft}.}
  \bibinfo{year}{2017}\natexlab{}.
\newblock \showarticletitle{Design of computer experiments: A review}.
\newblock \bibinfo{journal}{\emph{Computers \& Chemical Engineering}}
  \bibinfo{volume}{106} (\bibinfo{year}{2017}), \bibinfo{pages}{71--95}.
\newblock


\bibitem[\protect\citeauthoryear{Haghighi, Mehdipour, Bartocci, and
  Belta}{Haghighi et~al\mbox{.}}{2019}]%
        {haghighi2019control}
\bibfield{author}{\bibinfo{person}{Iman Haghighi}, \bibinfo{person}{Noushin
  Mehdipour}, \bibinfo{person}{Ezio Bartocci}, {and} \bibinfo{person}{Calin
  Belta}.} \bibinfo{year}{2019}\natexlab{}.
\newblock \showarticletitle{Control from signal temporal logic specifications
  with smooth cumulative quantitative semantics}. In
  \bibinfo{booktitle}{\emph{2019 IEEE 58th Conference on Decision and Control
  (CDC)}}. IEEE, \bibinfo{pages}{4361--4366}.
\newblock


\bibitem[\protect\citeauthoryear{Innes and Ramamoorthy}{Innes and
  Ramamoorthy}{2020}]%
        {innes2020probrobscene}
\bibfield{author}{\bibinfo{person}{Craig Innes} {and}
  \bibinfo{person}{Subramanian Ramamoorthy}.} \bibinfo{year}{2020}\natexlab{}.
\newblock \showarticletitle{ProbRobScene: A Probabilistic Specification
  Language for 3D Robotic Manipulation Environments}.
\newblock \bibinfo{journal}{\emph{arXiv preprint arXiv:2011.01126}}
  (\bibinfo{year}{2020}).
\newblock


\bibitem[\protect\citeauthoryear{Kulkarni, Kohli, Tenenbaum, and
  Mansinghka}{Kulkarni et~al\mbox{.}}{2015}]%
        {kulkarni2015picture}
\bibfield{author}{\bibinfo{person}{Tejas~D Kulkarni}, \bibinfo{person}{Pushmeet
  Kohli}, \bibinfo{person}{Joshua~B Tenenbaum}, {and} \bibinfo{person}{Vikash
  Mansinghka}.} \bibinfo{year}{2015}\natexlab{}.
\newblock \showarticletitle{Picture: A probabilistic programming language for
  scene perception}. In \bibinfo{booktitle}{\emph{Proceedings of the ieee
  conference on computer vision and pattern recognition}}.
  \bibinfo{pages}{4390--4399}.
\newblock


\bibitem[\protect\citeauthoryear{Mehdipour, Vasile, and Belta}{Mehdipour
  et~al\mbox{.}}{2019}]%
        {mehdipour2019arithmetic}
\bibfield{author}{\bibinfo{person}{Noushin Mehdipour},
  \bibinfo{person}{Cristian-Ioan Vasile}, {and} \bibinfo{person}{Calin Belta}.}
  \bibinfo{year}{2019}\natexlab{}.
\newblock \showarticletitle{Arithmetic-geometric mean robustness for control
  from signal temporal logic specifications}. In \bibinfo{booktitle}{\emph{2019
  American Control Conference (ACC)}}. IEEE, \bibinfo{pages}{1690--1695}.
\newblock


\bibitem[\protect\citeauthoryear{Niederreiter}{Niederreiter}{1992}]%
        {niederreiter1992random}
\bibfield{author}{\bibinfo{person}{Harald Niederreiter}.}
  \bibinfo{year}{1992}\natexlab{}.
\newblock \bibinfo{booktitle}{\emph{Random number generation and quasi-Monte
  Carlo methods}}.
\newblock \bibinfo{publisher}{SIAM}.
\newblock


\bibitem[\protect\citeauthoryear{Ritchie}{Ritchie}{2014}]%
        {ritchiequicksand}
\bibfield{author}{\bibinfo{person}{Daniel Ritchie}.}
  \bibinfo{year}{2014}\natexlab{}.
\newblock \showarticletitle{Quicksand: A Lightweight Embedding of Probabilistic
  Programming for Procedural Modeling and Design}. In
  \bibinfo{booktitle}{\emph{3rd NIPS Workshop on Probabilistic Programming}}.
  \bibinfo{pages}{164}.
\newblock


\bibitem[\protect\citeauthoryear{Tian, Fang, Tan, Qin, and Tang}{Tian
  et~al\mbox{.}}{2009}]%
        {tian2009uniform}
\bibfield{author}{\bibinfo{person}{Guo-Liang Tian}, \bibinfo{person}{Hong-Bin
  Fang}, \bibinfo{person}{Ming Tan}, \bibinfo{person}{Hong Qin}, {and}
  \bibinfo{person}{Man-Lai Tang}.} \bibinfo{year}{2009}\natexlab{}.
\newblock \showarticletitle{Uniform distributions in a class of convex
  polyhedrons with applications to drug combination studies}.
\newblock \bibinfo{journal}{\emph{Journal of multivariate analysis}}
  \bibinfo{volume}{100}, \bibinfo{number}{8} (\bibinfo{year}{2009}),
  \bibinfo{pages}{1854--1865}.
\newblock


\bibitem[\protect\citeauthoryear{Varnai and Dimarogonas}{Varnai and
  Dimarogonas}{2020}]%
        {varnai2020robustness}
\bibfield{author}{\bibinfo{person}{Peter Varnai} {and} \bibinfo{person}{Dimos~V
  Dimarogonas}.} \bibinfo{year}{2020}\natexlab{}.
\newblock \showarticletitle{On robustness metrics for learning STL tasks}. In
  \bibinfo{booktitle}{\emph{2020 American Control Conference (ACC)}}. IEEE,
  \bibinfo{pages}{5394--5399}.
\newblock


\bibitem[\protect\citeauthoryear{Zabinsky and Smith}{Zabinsky and
  Smith}{2013}]%
        {Zabinsky2013}
\bibfield{author}{\bibinfo{person}{Zelda~B. Zabinsky} {and}
  \bibinfo{person}{Robert~L. Smith}.} \bibinfo{year}{2013}\natexlab{}.
\newblock \bibinfo{booktitle}{\emph{Hit-and-Run Methods}}.
\newblock \bibinfo{publisher}{Springer US}, \bibinfo{address}{Boston, MA},
  \bibinfo{pages}{721--729}.
\newblock
\showISBNx{978-1-4419-1153-7}
\urldef\tempurl%
\url{https://doi.org/10.1007/978-1-4419-1153-7_1145}
\showDOI{\tempurl}


\bibitem[\protect\citeauthoryear{Zaremba}{Zaremba}{1966}]%
        {zaremba1966good}
\bibfield{author}{\bibinfo{person}{Stanis{\l}aw~K Zaremba}.}
  \bibinfo{year}{1966}\natexlab{}.
\newblock \showarticletitle{Good lattice points, discrepancy, and numerical
  integration}.
\newblock \bibinfo{journal}{\emph{Annali di matematica pura ed applicata}}
  \bibinfo{volume}{73}, \bibinfo{number}{1} (\bibinfo{year}{1966}),
  \bibinfo{pages}{293--317}.
\newblock


\bibitem[\protect\citeauthoryear{Zhang, Zhang, and Zhou}{Zhang
  et~al\mbox{.}}{2020}]%
        {zhang2020construction}
\bibfield{author}{\bibinfo{person}{Mei Zhang}, \bibinfo{person}{Aijun Zhang},
  {and} \bibinfo{person}{Yongdao Zhou}.} \bibinfo{year}{2020}\natexlab{}.
\newblock \showarticletitle{Construction of Uniform Designs on Arbitrary
  Domains by Inverse Rosenblatt Transformation}.
\newblock In \bibinfo{booktitle}{\emph{Contemporary Experimental Design,
  Multivariate Analysis and Data Mining}}. \bibinfo{publisher}{Springer},
  \bibinfo{pages}{111--126}.
\newblock


\end{thebibliography}

\end{document}